\documentclass[twocolumn,epjc3]{svjour3}

\usepackage{amsmath,bbm,amssymb}

\RequirePackage[T1]{fontenc}

\smartqed  

\RequirePackage{graphicx}
\RequirePackage{flushend}
\RequirePackage[numbers,sort&compress]{natbib}
\RequirePackage[colorlinks,citecolor=blue,urlcolor=blue,linkcolor=blue]{hyperref}

\journalname{Eur. Phys. J. C}

\begin{document}

\title{Constraining multi-Higgs flavour models}

\author{R. Gonz\'{a}lez Felipe\thanksref{e1,addr1,addr2} \and
    I. P. Ivanov\thanksref{e2,addr3,addr4,addr5} \and
    C. C. Nishi\thanksref{e3,addr6} \and
    Hugo Ser\^{o}dio\thanksref{e4,addr7} \and
    Jo\~{a}o P. Silva\thanksref{e5,addr1,addr2}}
\thankstext{e1}{e-mail: ricardo.felipe@ist.utl.pt}
\thankstext{e2}{e-mail: igor.ivanov@ulg.ac.be}
\thankstext{e3}{e-mail: celso.nishi@ufabc.edu.br}
\thankstext{e4}{e-mail: hugo.serodio@ific.uv.es}
\thankstext{e5}{e-mail: jpsilva@cftp.ist.utl.pt}
\institute{Instituto Superior de Engenharia de Lisboa - ISEL, Rua Conselheiro
    Em\'{\i}dio Navarro 1, 1959-007 Lisboa, Portugal\label{addr1} \and
    Centro de F\'{\i}sica Te\'{o}rica de Part\'{\i}culas (CFTP), Instituto Superior T\'{e}cnico,
    Universidade de Lisboa, Avenida Rovisco Pais 1, 1049-001 Lisboa,
    Portugal\label{addr2} \and
    IFPA, Universit\'{e} de Li\'{e}ge, All\'{e}e du 6 Ao\^{u}t 17, b\^{a}timent B5a, 4000 Li\'{e}ge,
    Belgium\label{addr3} \and
    Sobolev Institute of Mathematics, Koptyug avenue 4, 630090, Novosibirsk,
    Russia\label{addr4} \and
    Department of Physics and Astronomy, Ghent University, Proeftuinstraat 86,
    B-9000 Gent, Belgium\label{addr5} \and
    Universidade Federal do ABC - UFABC, Santo Andr\'{e}, SP, Brasil\label{addr6}
    \and
    Departament de F\'{\i}sica Te\`{o}rica and IFIC, Universitat de Val\`{e}ncia-CSIC, E-46100,
    Burjassot, Spain\label{addr7}}

\date{Received: date / Accepted: date}

\maketitle

\begin{abstract}
To study a flavour model with a non-minimal Higgs sector one must first define the symmetries of the fields; then identify what types of vacua exist and how they may break the symmetries; and finally determine whether the remnant symmetries are compatible with the experimental data. Here we address all these issues in the context of flavour models with any number of Higgs doublets. We stress the importance of analysing the Higgs vacuum expectation values that are pseudo-invariant under the generators of all subgroups. It is shown that the only way of obtaining a physical CKM mixing matrix and, simultaneously, non-degenerate and non-zero quark masses is requiring the vacuum expectation values of the Higgs fields to break completely the full flavour group, except possibly for some symmetry belonging to baryon number. The application of this technique to some illustrative examples, such as the flavour groups $\Delta(27)$, $A_4$ and $S_3$, is also presented.
\end{abstract}


\section{Introduction}
\label{sec:intro}

The origin of the flavour structure of the fermion masses and mixing in the standard model (SM) remains one of the unsolved puzzles in particle physics. Several approaches to this problem have been put forward, most of them based on the use of discrete~\cite{early_discrete} or continuous~\cite{Froggatt:1978nt} flavour symmetries. In particular, mainly motivated by the measurement of several neutrino oscillation parameters, the use of discrete symmetries has recently become more popular (for recent reviews, see e.g. Refs.~\cite{reviews_lepton}).

In flavour model building, one commonly chooses a flavour symmetry group $K$ and then studies how fermion masses and mixing are constrained by this symmetry. Because of the SM gauge symmetry there is, of course, an additional global hypercharge symmetry $U(1)_Y$ in the Lagrangian. Thus, in such studies it is important to distinguish the chosen flavour group $K$ from the full flavour group, denoted henceforth by $G=K\times U(1)_Y$. After spontaneous symmetry breaking (SSB) the vacuum expectation values (vevs) of the Higgs fields break completely, in many instances, $K$ and $G$, leaving no residual flavour symmetry in the model. But, in some rather interesting cases, the vevs could fully break $K$ but not $G$, and thus some residual symmetry remains. Clearly, any accidental symmetry of the Lagrangian, such as $U(1)_B$, corresponding to baryon number, is not relevant for the analysis of the flavour sector neither any residual symmetry of $G$ which is contained in it.

In the literature, there are studies focusing on symmetry groups exclusively
applied to the Higgs sector, either with
two~\cite{Ivanov:2007de,Nishi:2006tg,Maniatis:2007vn} or more than two
doublets~\cite{Ivanov:2012fp,Degee:2012sk,Keus:2013hya}, as well as studies
focusing either on flavour symmetry groups acting in the fermion sector with
assumed vev alignments~\cite{reviews_lepton,Felipe:2013ie,Felipe:2013vwa} or
on the residual symmetries remaining on the fermion mass
matrices~\cite{Leurer:1992wg,Lam:2008rs,Hernandez:2012ra}. Our aim is to
provide a contribution towards the connection of all these subjects. In
particular, we shall study in a systematic fashion the impact of a chosen
flavour group on the Higgs potential and its vevs, on the Yukawa coupling
matrices and thus on the residual physical properties of the mass matrices
and the Cabibbo-Kobayashi-Maskawa (CKM) quark mixing matrix.

The introduction of additional scalars is a common feature to all flavour models, which brings with it a large number of possible model implementations for a given flavour group. Studying all these implementations, even for two or three Higgs doublet models, is not an easy task, specially for large flavour groups. We shall show that most of the unphysical scenarios can be identified just by analysing how scalars transform under the flavour group and how they break it. Through this analysis we shall prove that under very general conditions some choices of flavour groups are excluded for a given scalar content. Although we shall restrict our discussion to the quark sector of the theory, the results presented here can be easily extended to the
lepton sector.

The paper is organised as follows. In Sect.~\ref{sec:theorem}, after briefly reviewing our notation and the experimental input, we analyse the invariance of the Yukawa coupling matrices under a full flavour group. Then we proceed to the formulation of a theorem which improves that in Ref.~\cite{Leurer:1992wg}, by including also the constraints arising from the right-handed quark sector. A summary of the possible CKM mixing patterns and quark spectra is presented, according to the existence or not of a residual symmetry in the whole quark sector. Most importantly, by introducing the notion of pseudo-invariance of the vev under all subgroups of the flavour group, we address for the first time the problem of knowing whether such a residual symmetry is possible or not. This issue has not been addressed before in the literature, and is explained in detail in Sect.~\ref{sec:examples}, where we illustrate the technique by applying it to the flavour groups $\Delta(27)$, $A_4$ and $S_3$, commonly found in the literature to explain fermion masses and mixing. Finally, our concluding remarks are given in Sect.~\ref{sec:conclusions}.

\section{A no-go theorem for models with $N$ Higgs doublets}
\label{sec:theorem}

\subsection{Notation and experimental input}

Let us consider a model with $N$ Higgs doublets, $\phi_k$, and three
generations of left-handed quark doublets, $Q_L = (u_L, d_L)^T $,
right-handed up-type quark singlets, $u_R$, and right-handed down-type quark
singlets, $d_R$. The Yukawa Lagrangian is
\begin{equation}
-{\cal L}_Y
=
\bar{Q}_L\, \Gamma_k\, \phi_k\, d_R
+
\bar{Q}_L\, \Delta_k\, \tilde{\phi}_k\, u_R
+
\textrm{H.c.},
\label{LY_before}
\end{equation}
where a sum over the Higgs fields ($k=1 \ldots N$) is implicit, $\Gamma_k$ and $\Delta_k$ are $3 \times 3$ matrices in the down- and up-type quark family space, respectively. The scalar fields can be combined into $\Phi = (\phi_1,\,\phi_2,\,\ldots,\,\phi_N)^T$. After the SSB, the scalars acquire vevs, $\langle \Phi \rangle = (v_1, \, v_2, \,
\ldots, \, v_N)^T$, and the Yukawa Lagrangian contains the terms
\begin{eqnarray}
- \langle {\cal L}_Y \rangle &=
&\left(
\bar{u}_L,  \bar{d}_L
\right)\, M_d\,
\left(
\begin{array}{c}
0\\
1
\end{array}
\right)\,
d_R
\nonumber\\
& &+\left(
\bar{u}_L,  \bar{d}_L
\right)\,
M_u\,
\left(
\begin{array}{c}
1\\
0
\end{array}
\right)\,
u_R
+
\textrm{H.c.},
\label{LY_after}
\end{eqnarray}
where
\begin{equation}
M_d = \Gamma_k\, v_k,\quad M_u = \Delta_k\, v_k^\ast, \label{MuMd_def}
\end{equation}
are the down-quark and up-quark mass matrices.

Using transformations that keep the $SU(2)_L$ doublet structure $\bar{Q}_L = \left( \bar{u}_L,  \bar{d}_L \right)$, it is possible to diagonalise $M_d$ or $M_u$, but not both matrices. We denote by ``left space'', the space of vectors $\bar{Q}_L$, which has three components in family space, one for each generation of quarks. So, $M_d$ and $M_u$ cannot be simultaneously diagonalised with left space transformations. To diagonalise them, we must treat $u_L$ and $d_L$ differently. Indeed, one can always find matrices $V_{uL}$, $V_{uR}$, $V_{dL}$, $V_{dR}$, such that
\begin{eqnarray}
V_{dL}^\dagger\, M_d\, V_{dR}
&=&
D_d = \textrm{diag}\,(m_d, m_s, m_b),
\nonumber\\
V_{uL}^\dagger\, M_u\, V_{uR}
&=&
D_u = \textrm{diag}\,(m_u, m_c, m_t).
\label{MdMu_diag}
\end{eqnarray}

Often, it is convenient to work with the Hermitian matrices
\begin{equation}
H_d = M_d\, M_d^\dagger\,, \quad H_u = M_u\, M_u^\dagger\,,
\label{HdHu_def}
\end{equation}
which live on the left space only and remain invariant under unitary
redefinitions of $u_R$ and/or $d_R$. Using Eqs.~\eqref{MdMu_diag} and
\eqref{HdHu_def}, we find
\begin{equation}
H_d = V_{dL}\, D_d^2\, V_{dL}^\dagger\,, \quad H_u = V_{uL}\, D_u^2\,
V_{uL}^\dagger\,.
\end{equation}

We know from experiment that the quark masses are non-vanishing and
non-degenerate. As a result, the matrices $H_d$ and $H_u$ are invertible and
have a non-degenerate spectrum. The CKM mixing matrix is given by
\begin{equation}
V = V_{uL}^\dagger V_{dL}.
\end{equation}
It is also known from experiment that $V$ must have three non-trivial angles
and that the largest contributor to CP violation in the kaon and $B$ meson
systems must be the CKM CP-violating phase, which is proportional to the
basis-invariant observable
\begin{equation}
J = \textrm{Det} \left( H_d H_u - H_u H_d\right).
\end{equation}
Therefore, $V$ cannot be block-diagonal or any permutation of a
block-diagonal matrix, it cannot be given by the identity or by some
permutation matrix, and $J$ must differ from zero.

\subsection{Yukawa coupling invariance under a full flavour group $G$}
\label{subsec:group}

Let us consider a group $G$ with group elements $g$. The group can be defined by a set of generators $i$. For example, we may have two generators, $a$ and $b$, such that all group elements are obtained by successive products of $a$ and $b$. Then one chooses how the various fields transform under the group $G$. This means that for three quark generations one chooses $3 \times 3$ matrices representing this group, so that the quark fields transform under the generators of the group as
\begin{eqnarray}
Q^G_{L\alpha}= \left(\mathcal{G}_{Li}\right)_{\alpha\beta}Q_{L\beta}\,,
\nonumber\\
d^G_{R\alpha}=
\left(\mathcal{G}^{d}_{Ri}\right)_{\alpha\beta}d_{R\beta}\,,\\
u^G_{R\alpha}= \left(\mathcal{G}^{u}_{Ri}\right)_{\alpha\beta}u_{R\beta}\,,\nonumber
\label{q_G}
\end{eqnarray}
where the index $i$ runs from $1$ to the number of group generators.
Similarly, one chooses $N \times N$ matrices, such that the scalar fields
transform as
\begin{equation} \Phi^G_{k}=
\left(\mathcal{G}_{i}\right)_{kl}\Phi_{l}\,. \label{phi_G}
\end{equation}
These equations hold for matrices $\mathcal{G}_i$ representing the generators as well as for matrices representing any element of the group, i.e. $\mathcal{G}_g$. The explicit form of $\mathcal{G}_{Li}$, $\mathcal{G}^{u}_{Ri}$, $\mathcal{G}^{d}_{Ri}$ and $\mathcal{G}_{i}$ dictates which representations $D(Q_L)$, $D(u_R)$, $D(d_R)$ and $D(\Phi)$, respectively, we have chosen for the fields.

We now ask the Yukawa Lagrangian to be invariant under this group before SSB. Using Eq.~\eqref{LY_before}, this implies the following relations for the couplings:
\begin{eqnarray}
\mathcal{G}_{Lg}^\dagger \,
\Gamma^k\,\mathcal{G}_{Rg}^d \,
\left(\mathcal{G}_g\right)_{kl}
=
\Gamma^l\,,
\nonumber\\
\mathcal{G}_{Lg}^\dagger \,
\Delta^k\,\mathcal{G}_{Rg}^u \,
\left(\mathcal{G}_g^\ast \right)_{kl}
=
\Delta^l\, ,
\label{inv}
\end{eqnarray}
for all group elements $g$. Equation~\eqref{inv} is clearly valid for the
identity element $g=e$. It is also valid for an element $g$ such that
$\mathcal{G}_{Lg}=\mathcal{G}_{Rg}=e^{i\theta}\mathbbm{1}$ and
$\mathcal{G}_g=\mathbbm{1}$. The latter relations are obviously always
verified since baryon number is conserved at or below the electroweak scale.
We call such an element a trivial group element, the group it generates
($G_g$) a trivial group, and we denote this case by $G_g\subseteq U(1)_B$.

To decide what invariance may remain in the Yukawa Lagrangian after SSB (and, therefore, in the mass matrices), we must know how the vacuum transforms under the group elements. The ensuing analysis may not be straightforward, so we will start with a simple case, and generalise it in steps. For the moment, we make three simplifying assumptions:
\begin{itemize}
\item[(i)] we ignore the up-type quarks;
\item[(ii)] we assume that $\Phi$ is in an irreducible representation
    (irrep) of $G$;
\item[(iii)] we assume that there is a non-trivial group element, $g_1$,
    leaving the vacuum invariant.
\end{itemize}

Given assumption (ii), we conclude that either no scalar couples to the
down-type quarks (which would lead to massless quarks) or else all scalars
couple to down-type quarks. This conclusion is a particular case of the
following more general assertion:
\newline\newline
\textbf{Proposition} \emph{For the down-type Yukawa terms in Eq.\,\eqref{LY_before}, each set of Higgs doublets $\phi_k$, comprising an
irreducible representation of $G$, either couples to quark fields with
linearly independent $\Gamma^k$, or decouples completely with $\Gamma^k=0$.}
\newline\newline
The same proposition holds for the up-type Yukawa terms and their coefficients $\Delta^k$. The proof is given in \ref{app:proof}; its essential idea is that if some $\Gamma^k$ were linearly dependent, then one would be able to identify doublets which do not couple to the down-quark fields and represent an invariant subspace under the action of $G$, this making the representation reducible. Thus, no matrix $\Gamma^k$ vanishes and
\begin{equation}
\left( \mathcal{G}_{g_1} \right)_{kl}\, v_l =
v_k,
\label{sym_vac}
\end{equation}
for some given group element $g_1$. The set of all elements $g_1$ satisfying
Eq.~\eqref{sym_vac} forms a subgroup of $G$, which we denote by $G_q$. The
Lagrangian after SSB is thus left invariant by a residual symmetry $G_q$.
Then, specializing in Eq.~\eqref{inv} to an element $g=g_1$ in $G_q$, and
using Eqs.~\eqref{MuMd_def} and~\eqref{sym_vac}, one gets
\begin{equation}
\mathcal{G}_{L{g_1}}^\dagger M_d\ \mathcal{G}_{R{g_1}}^d=M_d\,.
\label{symMd}
\end{equation}
Under these assumptions, $M_d$ is symmetric under the group element $g_1$ and the subgroup generated by it.

If assumption (ii) remains valid but we substitute (iii) by the assumption that the element $g \in U(1)_B$ is trivial (in the newly defined sense), then we conclude that $M_d$ has no symmetry, i.e., the group $G$ has been completely broken by the vacuum. In this case, we call the residual symmetry trivial, i.e. $G_q\subseteq U(1)_B$.

We now turn to the possibility that $\Phi$ is in a reducible representation
of $G$, thus negating our assumption (ii). First we assume that $\Phi$ breaks into the three vectors
\begin{equation}\label{reducible}
\left(\phi_1, \ldots, \phi_r \right),
\ \ \
\left(\phi_{r+1}, \ldots, \phi_s \right),
\ \ \
\left(\phi_{s+1}, \ldots, \phi_N \right),
\end{equation}
which transform as the irreducible representations $\varphi_1$, $\varphi_2$, and $\varphi_3$ of $G$, respectively. We can still write Eq.~\eqref{inv}. The only novelty is that $\mathcal{G}_g$ is now a block-diagonal matrix, with one $r \times r$ block, one $(s-r) \times (s-r)$ block, and one $(N-s) \times (N-s)$ block. Said otherwise, we may write
\begin{eqnarray}
\mathcal{G}_{Lg}^\dagger \,
\Gamma^k_{\varphi_1}\, \mathcal{G}_{Rg}^d \,
\left( \mathcal{G}_g^{\varphi_1} \right)_{kl}
&=&
\Gamma^l_{\varphi_1},
\ \
k,l = 1 \dots r,
\label{sym_G_vp1}
\\
\mathcal{G}_{Lg}^\dagger \,
\Gamma^k_{\varphi_2}\, \mathcal{G}_{Rg}^d \,
\left( \mathcal{G}_g^{\varphi_2} \right)_{kl}
&=&
\Gamma^l_{\varphi_2},
\ \
k,l = r+1 \dots s,
\label{sym_G_vp2}
\\
\mathcal{G}_{Lg}^\dagger \,
\Gamma^k_{\varphi_3}\, \mathcal{G}_{Rg}^d \,
\left( \mathcal{G}_g^{\varphi_3} \right)_{kl}
&=&
\Gamma^l_{\varphi_3},
\ \
k,l = s+1 \dots N,
\label{sym_G_vp3}
\end{eqnarray}
where we introduce a label $\varphi_j$ denoting the representation. We assume that the $\varphi_3$ representation is such that the fields $\left(\phi_{s+1}, \ldots, \phi_N \right)$ do not couple to the down-type quarks, i.e., $\Gamma^k_{\varphi_3} = 0$. In this case,
\begin{eqnarray}
M_d =
\sum_{k=1}^N \Gamma^k v_k
=
\sum_{k=1}^r \Gamma^k_{\varphi_1} v_k^{\varphi_1}
+
\sum_{k=r+1}^s \Gamma^k_{\varphi_2} v_k^{\varphi_2}.
\label{Md_vp1_vp2}
\end{eqnarray}
We further assume that the vevs of $\left(\phi_{1}, \ldots, \phi_r \right)$
and $\left(\phi_{r+1}, \ldots, \phi_s \right)$ are invariant under some
subgroups $G_1$ and $G_2$, respectively. Thus,
\begin{equation}
\left(\mathcal{G}_{g_1}\right)_{kl}\, v_l^{\varphi_1}
=
v_k^{\varphi_1},
\quad
\left(\mathcal{G}_{g_2}\right)_{kl}\, v_l^{\varphi_2}
=
v_k^{\varphi_2},
\label{sym_vac_vp2}
\end{equation}
for $g_1\in G_1$ and $g_2 \in G_2$. The common residual symmetry is now $G_q=G_1\cap G_2$. If $G_q\subseteq U(1)_B$, then the residual symmetry is trivial. In contrast, if we have a non-trivial $G_q\not\subseteq U(1)_B$, we can pick $g\in G_q$ and using Eqs.~\eqref{sym_G_vp1} and \eqref{sym_G_vp2}, together with Eqs.~\eqref{Md_vp1_vp2} and \eqref{sym_vac_vp2}, we obtain again Eq.~\eqref{symMd}. Notice that, in this example, $\varphi_3$ is irrelevant for $M_d$ but it may matter for $M_u$, should $\left(\phi_{s+1}, \ldots, \phi_N \right)$ couple to the up-type quarks.

We are now ready to generalise the impact of symmetries and to include the up-quark sector. The residual symmetry $G_q$ restricted to the down-quark sector is now renamed $G_d$, while $G_q$ is reserved for the residual symmetry of the quark sector as a whole. If $\Phi$ is in a reducible representation of $G$, it may be seen as a sum of irreducible representations $\varphi_i$ as
\begin{equation}
\Phi = \varphi_1 \oplus \varphi_2 \oplus
\cdots
\oplus \varphi_m
\quad \textrm{with}
\quad
m \leq N.
\label{Phi_decomp}
\end{equation}
Equation~\eqref{inv} can now be written as a set of equations for each
$\varphi_j$,
\begin{eqnarray}
\mathcal{G}_{L{g}}^\dagger \,
\Gamma^k_{\varphi_j}\,\mathcal{G}_{R{g}}^d \,
\left( \mathcal{G}^{\varphi_j}_{g} \right)_{kl}
=
\Gamma^l_{\varphi_j}\,,
\nonumber\\
\mathcal{G}_{L{g}}^\dagger \,
\Delta^k_{\varphi_j}\,\mathcal{G}_{R{g}}^u \,
\left( \mathcal{G}_{g}^{\varphi_j} \right)^\ast _{kl}
=
\Delta^l_{\varphi_j}\,,
\label{inv2}
\end{eqnarray}
where no sum is assumed in $\varphi_j$. The matrices $\Gamma^k_{\varphi_j}$ and $\Delta^k_{\varphi_j}$ denote the couplings associated with $\phi_k$ for a given irreducible representation $\varphi_j$. For example, if $\Phi$ is an irreducible quadruplet $(\phi_1,\phi_2,\phi_3,\phi_4)$ then $\Gamma^k$, for $k=1 \dots 4$, denotes the Yukawa coupling associated with each entry of the scalar representation. If $\Phi$ is a reducible quadruplet, composed by two irreducible doublets $\varphi_1 = (\phi_1,\phi_2)$ and $\varphi_2 = (\phi_3,\phi_4)$, then the coupling $\Gamma_{\varphi_i}^k$ associates two couplings to each irreducible representation, i.e. $k=1,2$. One denotes the vev of the irreducible representations $\varphi_j$ as $v^{\varphi_j}$. When the vev breaks the group $G$, the Lagrangian may still remain invariant under the action of some elements of $G$.

Let us denote by $G_d$ and $G_u$ the subgroups of $G$ that are left invariant by the vevs of all Higgs fields coupling to down-type and up-type quarks, respectively. Some fields may interact with both sectors. The relevant residual symmetry of the quark sector is $G_q=G_d\cap G_u$. The fields that do not couple to quarks are irrelevant to our discussion. There are several possibilities:
\begin{enumerate}
\item[(1)] If $G_q\subseteq U(1)_B$, there is no non-trivial symmetry left
    in the quark sector as a whole. More specifically,
\begin{enumerate}
\item If $G_{d,u} \subseteq U(1)_B $, then there is no non-trivial
    symmetry left in either (up- or down-) quark sector;
\item If $G_d \not\subseteq U(1)_B$ and $G_u\subseteq U(1)_B$, then there
    is some non-trivial symmetry left in $M_d$, but not in $M_u$;
\item If $G_d \subseteq U(1)_B$ and $G_u \not\subseteq U(1)_B$, then
    there is some non-trivial symmetry left in $M_u$, but not in $M_d$;
\item If $G_d \not\subseteq U(1)_B$ and $G_u \not\subseteq U(1)_B$, then
    there is some non-trivial symmetry left in $M_d$, some non-trivial
    symmetry left in $M_u$, but there is no non-trivial symmetry left in
    both $M_d$ and $M_u$;
\end{enumerate}

\item[(2)] If $G_q\not\subseteq U(1)_B$, then there is some non-trivial
    symmetry left in the quark sector as a whole.

\end{enumerate}

The characteristic in common to the cases in (1) is that $\langle \Phi
\rangle$ breaks the symmetry completely, modulo $U(1)_B$. In contrast, case
(2) leaves a common non-trivial symmetry in the whole quark sector. In the
latter case, for $g \in G_q$, from Eqs.~\eqref{inv2} we find
\begin{equation}
\mathcal{G}_{L{g}}^\dagger \,
M_d\
\mathcal{G}_{R{g}}^d \,
=
M_d ,
\quad
\mathcal{G}_{L{g}}^\dagger \,
M_u\
\mathcal{G}_{R{g}}^u \,
=
M_u .
\label{inv2MuMd}
\end{equation}
with $M_d = \Gamma^k_{\varphi_j} v^{\varphi_j}_k$ and $M_u = \Delta^k_{\varphi_j} v_k^{\varphi_j\, \ast}$. Equation~\eqref{inv2MuMd} generalises Eq.~\eqref{symMd} to the case of reducible representations and is the main result of this section. Thus, when using Eq.~\eqref{inv2MuMd}, as we will do in the theorem to be proved in the next section, it does not matter whether the representation of the group $G$ is reducible or irreducible.

\subsection{Theorem and proof}
\label{subsec:theorem}

Let us consider a model with quarks and scalar fields transforming under some set of representations of a full flavour group $G$. We denote the representation space where $G$ acts as the \textit{flavour space}, i.e., the horizontal space of replicated multiplets of quark and scalar fields with the same gauge quantum numbers. The following theorem strongly constrains the viable models.
\newline\newline
\textbf{Theorem} \emph{(No-Go) Given a group $G$ acting on the fla\-vour space, the only way to obtain a non-block-diagonal CKM mixing matrix and, simultaneously, non-de\-gen\-er\-ate and non-zero quark masses, is that $\left<\Phi\right>$ breaks completely the group $G$, except possibly for some symmetry belonging to baryon number.}
\newline\newline
\noindent\emph{Proof} Suppose that there is a residual symmetry group $G_q$ which is left invariant by the vev $\left<\Phi\right>$. Let us denote a generic element of $G_q$ as $g_1$, and then Eq.~\eqref{sym_vac} is assumed.\footnote{Here we only consider Higgs doublets that couple to quarks.} We want to show that either $g_1=e$, i.e., the trivial element in $G$, or $g_1$ acts on the flavour space as a member of the baryon number symmetry.

Equation~\eqref{sym_vac} implies that the Lagrangian after SSB is invariant
under $g_1$. Using the same rationale that led to Eq.~\eqref{inv2MuMd}, we
conclude that
\begin{equation}
\mathcal{G}_{L{g_1}}^\dagger M_d\ \mathcal{G}_{R{g_1}}^d
=
M_d\,,
\quad
\mathcal{G}_{L{g_1}}^\dagger M_u\ \mathcal{G}_{R{g_1}}^u
=
M_u\,.
\label{symMuMd}
\end{equation}
Since we shall be interested in the left sector, where the CKM mixing arises, we may turn to the Hermitian combinations in Eq.~\eqref{HdHu_def}, which verify
\begin{equation}
\mathcal{G}_{L{g_1}}^\dagger H_d\ \mathcal{G}_{L{g_1}}
=
H_d\,,
\quad
\mathcal{G}_{L{g_1}}^\dagger H_u\ \mathcal{G}_{L{g_1}}
=
H_u\,.
\label{symHuHd}
\end{equation}

The next step is to use the following result (proved in \ref{block.diag}): if the generator $\mathcal{G}_{Lg_1}$ given in Eq.~\eqref{symHuHd} is not proportional to the identity matrix, then the CKM matrix is block-diagonal, in contradiction with experiment. Note that we include any permutation of a block-diagonal matrix, as well as the unit mixing matrix, $V=\mathbbm{1}$, in the category of block-diagonal matrices. We then conclude that
\begin{equation}
\mathcal{G}_{L{g_1}} =
e^{i \theta}\, \mathbbm{1},
\label{H_GL_3}
\end{equation}
so that Eqs.~\eqref{symHuHd} place no restriction on $H_{d,u}$, and one has a potentially viable CKM matrix.

Returning now to Eqs.~\eqref{symMuMd}, it follows that
\begin{equation}
\mathcal{G}_{R{g_1}}^d = \mathcal{G}_{R{g_1}}^u = \mathcal{G}_{L{g_1}},
\label{GR=GL}
\end{equation}
because $M_u$ and $M_d$ are invertible. Applying this to Eq.~\eqref{inv}, we
find
\begin{equation}
\Gamma^k\, \left( \mathcal{G}_{g_1} -\mathbbm{1}\right)_{kl}=\boldsymbol{0},
\quad
\Delta^k\, {\left( \mathcal{G}_{g_1}-\mathbbm{1} \right)^\ast}_{kl}=\boldsymbol{0},
\label{GamDel_g1}
\end{equation}
for $l=1,\ldots,N$. This implies that
\begin{equation}
\mathcal{G}_{g_1}= \mathbbm{1},
\end{equation}
within each space where the set of Yukawa matrices $\{\Gamma^k\}$ (or,
analogously, $\{\Delta^k\}$) is linearly independent. Since we are confined
to Higgs doublets that couple to quarks -- down-type, up-type or both -- the
proposition in Sect.~\ref{subsec:group} implies that one or both of the
sets $\{\Gamma^k\}$ and $\{\Delta^k\}$ are composed of non-zero and linearly
independent matrices for each irreducible sector of $\Phi$. Hence,
\begin{equation}
\mathcal{G}_{g_1} = \mathbbm{1}_{n \times n},
\label{is1}
\end{equation}
in the whole space of Higgs doublets that couple to quarks.

Finally, we can analyse Eqs.~\eqref{H_GL_3}, \eqref{GR=GL} and \eqref{is1} jointly. If $\Phi$ is in a faithful representation, then $g_1=e$ and no residual symmetry is present. In this case, $e^{i\theta}=1$ in Eq.~\eqref{H_GL_3}. If $g_1\neq e$ ($e^{i\theta}\neq 1$), then $\Phi$ is unfaithful and some residual symmetry will be present in the final Lagrangian, without constraining the mixing. This residual symmetry should be specifically represented by Eqs.~\eqref{H_GL_3}, \eqref{GR=GL} and \eqref{is1}, which is just part of the baryon number conservation. This completes the proof of the theorem, which generalises that in Ref.~\cite{Leurer:1992wg}, by including the constraints on the right-handed quark sector. Without the latter constraints, theories leading to unphysical massless quarks would not be precluded.\footnote{We give one such example at the end of Sect.~\ref{sec:A4}.}

At this point two remarks are in order. First, any SM-like Lagrangian, as the one in Eq.~\eqref{LY_before}, exhibits automatic conservation of the baryon number $U(1)_B$, independently of additional gauge or flavour symmetries. Such accidental symmetry imposes no constraints on masses and mixing, and remains conserved after electroweak symmetry breaking. The proof above shows that a non-block-diagonal $V_{CKM}$ is compatible with some residual symmetry inside $G$ only if the latter is a subgroup belonging to $U(1)_B$. Second, we should stress that Eq.~\eqref{is1} applies only to the Higgs doublets that couple to quarks. The Higgs doublets that appear solely in the scalar potential are important when analysing possible vacuum alignments, but they are irrelevant for the statement of the theorem.

Notice that no information of the scalar potential and vevs has been used in
this proof. That is, the theorem constrains residual symmetries and can be
applied to virtually any model. Nevertheless, our interest in this article
lies on NHDM, and, as we will show in Sect.~\ref{sec:examples}, the
explicit transformation properties of the Higgs scalars, their vevs, and
their relation to the subgroup chains are crucial, in particular, for the
application of the theorem to specific symmetries of complete models of
scalars and fermions.

\subsection{Some illustrative examples}
\label{subsec:examples}

The theorem proved in the previous section is completely general; yet it is
useful to study in more detail several particular cases that one may come
across. We split them into four classes depending on whether the
representations $D(Q_L)$  and $D(\Phi)$ of $G$, acting, respectively, on the
left quark space $Q_L$ and the scalar space $\Phi$, are faithful or not. We
denote the kernel of the representation of $D(Q_L)$ by $\ker D(Q_L)$ and
write $\ker D(Q_L)=\{e\}$ if the representation is faithful. Otherwise $\ker
D(Q_L)$ is the subgroup of $G$ which is mapped to the identity by the
representation of $Q_L$.

\begin{enumerate}
\item[(i)] $Q_L$ and $\Phi$ faithful: $\ker D(Q_L) =\ker D(\Phi)=\{e\}$

A direct application of the theorem implies that if $\left<\Phi\right>$
does not break $G$ completely, then there is a residual symmetry in the
fermion sector leading to a block-diagonal mixing.

\item[(ii)] $Q_L$ unfaithful and $\Phi$ faithful: $\ker D(Q_L)\neq \{e\}$
    and $\ker D(\Phi)=\{e\}$

The full symmetry in the left sector is smaller than the one in the scalar
sector and, in principle, we could have the proper subgroup $\ker D(Q_L)$
of $G$ (or smaller) unbroken by $\left<\Phi\right>$. However,
Eq.~\eqref{is1} should hold and $\left<\Phi\right>$ should break $G$
completely.

As an example, let us take $G=A_4$ with $Q_L\sim
(\boldsymbol{1},\boldsymbol{1}',\boldsymbol{1}'')$ and $\Phi\sim d_R\sim
u_R\sim \boldsymbol{3}$. If $\left<\Phi\right>\sim(1,0,0)$, the
$\mathbb{Z}_2$ subgroup generated by $g_1=\mathrm{diag}\,(1,-1,-1)$ is
conserved and is contained in $\ker D(Q_L)$. The representation
$\mathcal{G}_{Lg_1}=\mathbbm{1}_3$ satisfies Eq.~\eqref{H_GL_3}, but
Eqs.~\eqref{GR=GL} and \eqref{is1} are not valid as
$\mathcal{G}_{g_1}=\mathcal{G}_{Rg_1}^d=\mathcal{G}_{Rg_1}^u=\mathrm{diag}\,(1,-1,-1)$.
Therefore, we end up with a non-trivial residual symmetry, but the
invertibility of $M_u$ and $M_d$ assumed in Eq.~\eqref{GR=GL} is lost.

\item[(iii)] $Q_L$ faithful and $\Phi$ unfaithful: $\ker D(Q_L)=\{e\}$ and
    $\ker D(\Phi) \neq\{e\}$

This case is automatically excluded unless $\ker D(\Phi)$ acts like baryon number on quarks. The full symmetry of the potential is $G/\ker D(\Phi)$ and $\left<\Phi\right>$ can never break $\ker D(\Phi)$. As an example, let us take $G=A_4$ with $(\phi_1,\phi_2,\phi_3)\sim (\boldsymbol{1},\boldsymbol{1}',\boldsymbol{1}'')$ and $Q_L\sim d_R\sim u_R\sim \boldsymbol{3}$. If all $\phi_k$ get non-zero vevs, they only break $A_4$ to $\mathbb{Z}_2\times\mathbb{Z}_2$, which corresponds to the kernel of the representation of $\Phi$, and is generated by $\mathrm{diag}\,(1,-1,-1)$ and $\mathrm{diag}\,(-1,-1,1)$. This subgroup remains in the $Q_L$ quark sector and the CKM matrix would be trivial.

\item[(iv)] $Q_L$ and $\Phi$ unfaithful: $\ker D(Q_L) \neq \{e\}$ and $\ker
    D(\Phi)$ $\neq \{e\}$

This case can be discarded given the Yukawa structure in
Eq.~\eqref{LY_before} and the assumption that at least one of the
representations for $Q_L,\Phi,d_R,u_R$ is a faithful irrep or contains a
faithful irrep (i.e. $G$, and not a smaller group, is the full flavour
symmetry); see also \ref{ap:yukawa}.

\end{enumerate}

During the proof and consequent remarks of the theorem, nothing has been said about the reducibility or irreducibility of $\Phi$. If $\Phi$ is an irreducible representation of the full flavour group $G$, then the same field components $\phi_i$ couple to the up and down sectors. In the case that $\Phi$ is a reducible representation of $G$, the components $\phi_i$ can be arranged into irreducible multiplets just like in Eq.~\eqref{Phi_decomp}. In this case, there are many ways of breaking the flavour group $G$ completely in order to allow for models with non-zero and non-degenerate masses with viable $V_{CKM}$ (see end of Sect.~\ref{subsec:group} for the different possibilities).

We shall comment on case (1d) of Sect.~\ref{subsec:group}, which is the most predictive and hence one of the most common approaches in model building, since each quark sector has its own non-trivial residual symmetry while the full flavour group $G$ is completely broken. This will lead, in general, to strong constraints to the mixing angles and the CP phase. For instance, in Ref.~\cite{Araki:2013rkf}, the authors studied which would be the best residual symmetries in order to accommodate the experimental CKM mixing matrix. This was done in the context of the von Dyck flavour groups~\cite{Hernandez:2012ra}. It is shown that the flavour groups $D_N$ and $\Delta(6N^2)$, with $N$ an integer multiple of 7, are capable of determining the mixing angle $\theta_{12}$ close to its experimental value, if we fix $\theta_{23}=\theta_{13}=0$. Non-zero values for all three mixing angles require, on the other hand, infinite von Dyck groups.

We summarise in Table~\ref{table:constraints} the different possibilities for the CKM mixing angles and mass spectra depending on the existence (or not) of a residual symmetry in the whole theory. In the presence of a residual symmetry, not contained in $U(1)_B$, we can either have at most three non-zero mixing angles with some massless quarks, as in example (ii) above, or have massive quarks with at most one non-zero mixing angle, as in example (iii).

\begin{table*}
\caption{Possible CKM mixing patterns and quark spectra depending on the presence of global residual symmetries, modulo $U(1)_B$. The Yes/No distinguishes if one/two quarks of the same type are massless.}
\label{table:constraints}
\begin{tabular*}{\textwidth}{@{\extracolsep{\fill}}cccc@{}}
\hline
\,Residual symmetry\, &\, Non-zero mixing angles\, & \,CP violation\, &\, Quark spectrum\,\\
\hline
Yes&$\leq 1$&No&All masses $\neq 0$\\
Yes&$\leq 3$&Yes/No&Some masses $= 0$\\
No&$\leq 3$&Yes&All masses $\neq 0$\\
\hline
\end{tabular*}
\end{table*}

\section{Application of the theorem to flavour models}
\label{sec:examples}

The no-go theorem previously demonstrated strongly constrains viable flavour models based on a full flavour symmetry $G$ of the Lagrangian. However, it is crucial for its application that one can determine the symmetry properties of the vevs. In this section, we shall perform this task by introducing the notion of pseudo-invariance of the vevs under subgroups of the original group. We illustrate the technique by applying it to some examples of flavour groups commonly found in the literature to explain fermion masses and mixing.

Before proceeding with the examples, it is worth recalling the distinction between what is meant by ``full flavour group'' and by simply ``flavour group''. As said before, in model building, we may add symmetries that act only on the flavour space of the model. Those symmetries are usually taken to be subgroups of $SU(N)$ (in particular $SU(3)$, when we are dealing with three fermion generations). We refer to these symmetries as flavour symmetries $K$. However, the full flavour symmetry group $G$ is, in general, larger since it contains the global hypercharge transformation, i.e. $G=K\times U(1)_Y$.\footnote{The center of $SU(N)$ is $Z(SU(N))=\mathbb{Z}_N$, which is already included in the $U(1)$ global transformations of $G$. To avoid redundancy we may work with the projective group $PSU(N)=SU(N)/\mathbb{Z}_N$, and define the flavour symmetry as a subgroup of $PSU(N)$ instead.} The accidental baryon symmetry $U(1)_B$ is not included in the full flavour symmetry $G$ because the Higgs fields do not transform under it, but it might happen that $G$ intersects $U(1)_B$. We further assume that there is no additional accidental symmetry in the Yukawa Lagrangian \eqref{LY_before}, apart from $G$ and $U(1)_B$ (see Refs.\,\cite{yukawa:abelian} for methods for detecting them).

The additional $U(1)_Y$ in the full flavour symmetry is of extreme importance. As we shall see, one may break completely the flavour symmetry $K$ but still leave a non-trivial residual symmetry in $G$. To better understand this scenario, we introduce the notion of pseudo-invariant vacuum. In our discussion we have used the invariance of the vacuum, i.e. Eq.~\eqref{sym_vac}, in order to constrain the possible viable models. This vacuum invariance has been defined in the full flavour group $G$. Looking only to the flavour group $K$, we may extend the notion of invariance to pseudo-invariance of an irrep through the condition
\begin{equation} \left( \mathcal{G}_{\tilde{g}_1} \right)_{kl}\, v_l =
e^{i\alpha}v_k\quad\text{with}\quad \tilde{g}_1\in K. \label{pseudsym_vac}
\end{equation}
The group element $\tilde{g}_1$ changes the vacuum by a global phase transformation. In general, these global phase transformations are not contained in $K$, but they are contained in $G$ due to the $U(1)_Y$ global group. This means that, for a given irrep, any pseudo-invariant vacuum in $K$ can be written as an invariant one in $G$. Therefore, instead of dealing with the full flavour group $G$, we may restrict ourselves to the flavour group $K$ and its non-trivial pseudo-invariant vevs. Note that if $\Phi$ is composed of more than one irrep $\varphi_j$, as in Eq.\,\eqref{Phi_decomp}, a pseudo-invariant vev where
\begin{equation}
\left( \mathcal{G}_{\tilde{g}_1} \right)_{kl}\, v_l^{\varphi_j} =
e^{i\alpha_j}v_k^{\varphi_j}\,,\quad\text{with}\quad \tilde{g}_1\in K\,,
\label{pseudsym_vac:irreps}
\end{equation}
is equivalent to an invariant vev of $G$ only if all vevs in each irrep transform by the same global phase, i.e., $\alpha_j=\alpha$. One consequence immediately follows: the presence of an invariant Higgs doublet, singlet with respect to $K$, may be relevant. For example, consider down-type quarks interacting only with one $\mathbf{n}$-dimensional irrep $\varphi_1=(\phi_1,\ldots,\phi_n)$ whose vev is pseudo-invariant by an element $\tilde{g}_1 \neq e$ with $\alpha_1\neq 0$. The mass matrix $M_d$ will possess a non-trivial residual symmetry. Now, suppose we add a trivial irrep $\varphi_2=\phi_{n+1}\sim\mathbf{1}$ that also interacts with down-type quarks. Any vev for $\varphi_2$ will be strictly invariant by $K$, with $\alpha_2=0$ in Eq.\,\eqref{pseudsym_vac:irreps}. Hence, $M_d$ will no longer possess the previous residual symmetry.

We remark that Eq.~\eqref{pseudsym_vac:irreps} (and consequently Eq.~\eqref{pseudsym_vac}) is merely an eigenvalue equation. Therefore, from a simple group-theoretical method, and without analysing the scalar potential, we can extract a set of vacuum alignments that will be automatically excluded by the theorem, namely those corresponding to the eigenvectors of $\mathcal{G}_{\tilde{g}_1}$. Notice, however, that the vev alignments obtained through this procedure may not be global minima of the scalar potential. Yet, this is a straightforward group-theoretical check in the spirit familiar to model building. One could instead follow a geometrical method~\cite{Degee:2012sk} to find first the global minima of the scalar potential. In the latter case, if all the minima preserve some subgroup of the initial symmetry, the theorem applies directly without the need of solving the eigenvalue equation~\eqref{pseudsym_vac:irreps}. The drawback is that the geometrical approach is not universal, and it is not guaranteed that it could easily be applied to more complicated Higgs sectors, for example to some high symmetry groups in 4HDM.

Next we shall present three examples of flavour groups in the context of 3HDM, two of them with $\Phi$ being in a faithful triplet representation, for the $\Delta(27)$ and $A_4$ flavour groups, and one example with $\Phi$ in a reducible triplet representation, for the $S_3$ flavour group.

\begin{figure*}[t]
\centering
\includegraphics[width=15cm]{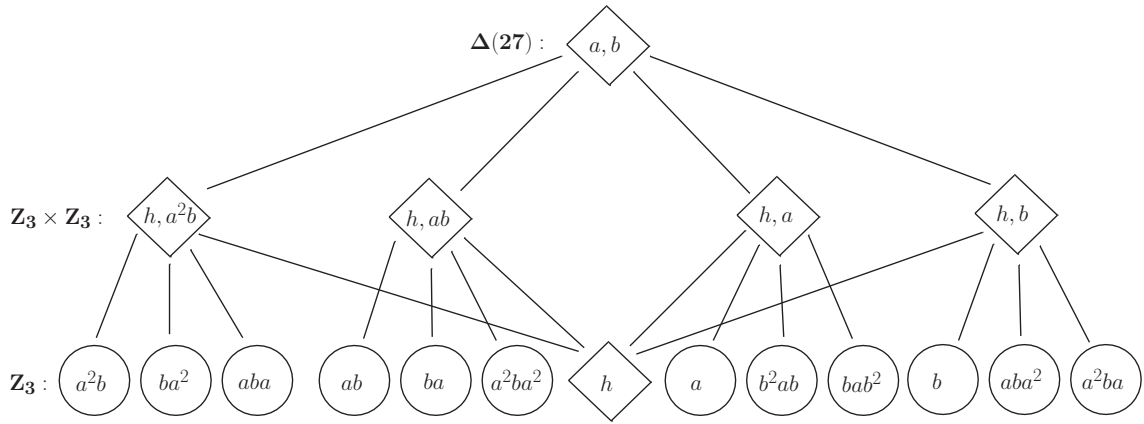}
\caption{Diagram of the subgroups of $\Delta(27)$ and their generators, based on Ref.~\cite{hobbes}. Invariant subgroups are encased in diamond shaped boxes; others are in circles.}
\label{fig:Delta27_graph}
\end{figure*}

\subsection{The flavour group $\Delta(27)$ in 3HDM}
\label{sec:D27}

The flavour group $\Delta(27)$~\cite{Luhn:2007uq} is a subgroup of $SU(3)$
and can be viewed as the group defined by two generators $a$ and $b$, with
presentation
\begin{equation}
a^3 = b^3 = (a b)^3 = 1.
\label{presentation}
\end{equation}
We use the specific three-dimensional implementation~\cite{Ivanov:2012fp}:
\begin{equation} a = \left(
\begin{array}{ccc}
1 & 0 & 0\\
0 & \omega & 0\\
0 & 0 & \omega^2
\end{array}
\right), \ \ \ b = \left(
\begin{array}{ccc}
0 & 1 & 0\\
0 & 0 & 1\\
1 & 0 & 0
\end{array}
\right), \label{D27:a_and_b} \end{equation}
where $\omega = e^{2 i \pi/3}$. The group has 27 elements divided into 11
irreducible representations: nine singlets $\mathbf{1}_{(i,j)}$, with $i,j =
0,1,2$; and two triplets, $\mathbf{3}_{(0,1)}$ and $\mathbf{3}_{(0,2)}$.
There are two elements of particular interest. One is
\begin{equation} d \equiv b^2\, a\, b = \left(
\begin{array}{ccc}
\omega^2 & 0 & 0\\
0 & 1 & 0\\
0 & 0 & \omega
\end{array}
\right). \label{D27:d} \end{equation}
Indeed, since $\Delta(27)$ is isomorphic to $Z_3 \times Z_3 \rtimes Z_3$ we
can view the first two $Z_3$'s as generated by $a$ and $d$ (they obviously
commute), while the third one is generated by $b$. The second element of
interest is
\begin{equation}
h \equiv a^2\, b\, a\, b^2 = \omega\, \left(
\begin{array}{ccc}
1 & 0 & 0\\
0 & 1 & 0\\
0 & 0 & 1
\end{array}
\right).
\label{D27:h}
\end{equation}
The $Z_3$ group generated by $h$ is the only $Z_3$ invariant subgroup of $\Delta(27)$. A summary of the subgroups and their generators is presented in Fig.~\ref{fig:Delta27_graph}, based on Ref.~\cite{hobbes}. The eigenvectors $(v_1, v_2, v_3)$ associated to the generators of the $Z_3 \times Z_3$ invariant subgroups of $\Delta(27)$ are given by
\begin{eqnarray}
\begin{split}
a^2b:&&\  (1, 1,
\omega),\ \ (1, \omega, 1),\ \
 (\omega, 1, 1);\\
 ab:&&\  (1, 1, \omega^2),\ \
(1, \omega^2, 1),\ \
 (\omega^2, 1, 1);\\
  a:&&\  (0, 0, 1),\ \
(0,1, 0),\ \
 (1, 0, 0);\\
  b:&&\ (1, 1, 1),\ \ (1,\omega, \omega^2),\ \ (1,\omega^2, \omega).
\end{split}
\label{nvevD27}
\end{eqnarray}

\begin{figure*}[t]
\centering
\includegraphics[width=11cm]{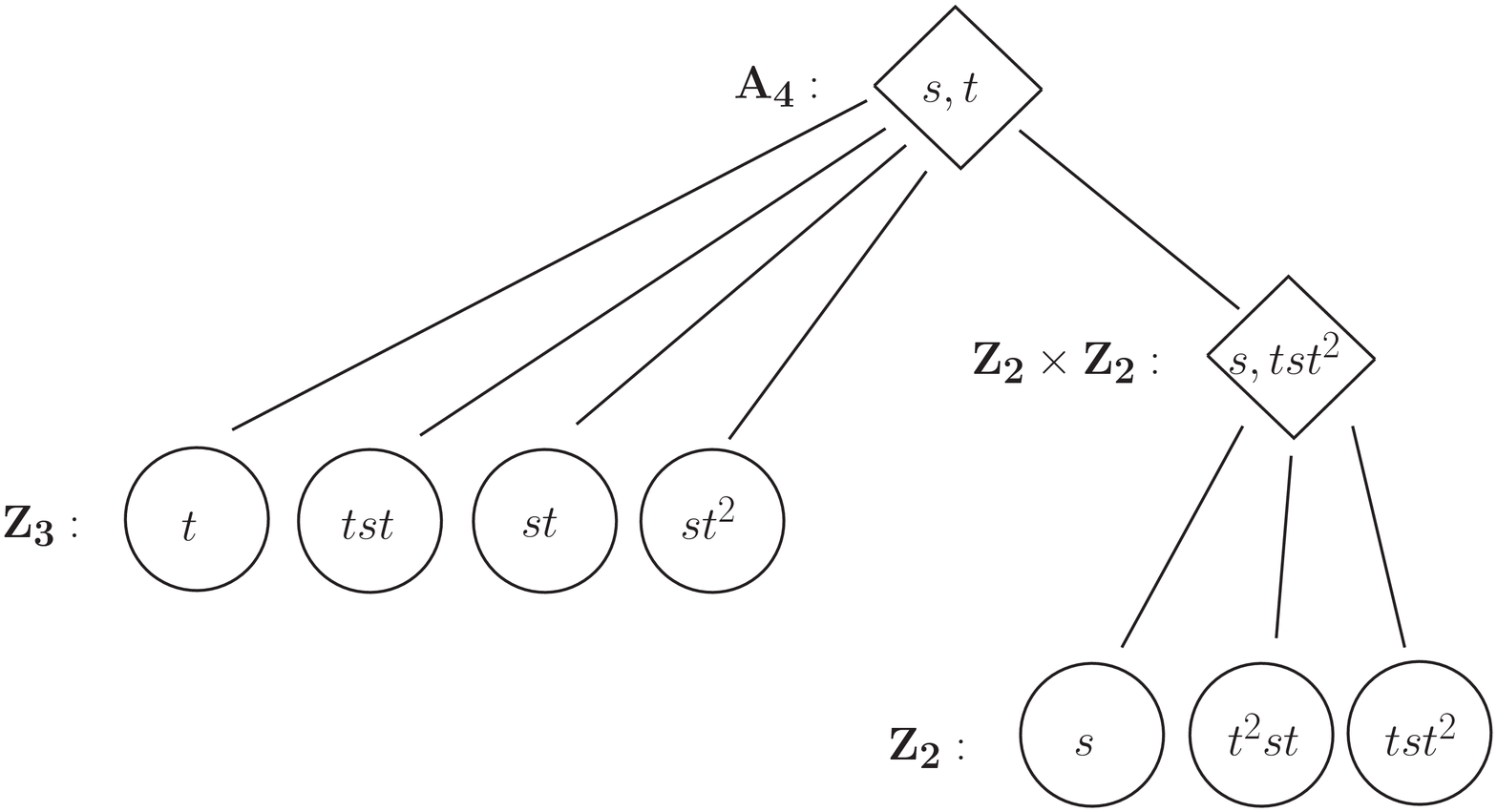}
\caption{Diagram of the subgroups of $A_4$ and their generators, based on Ref.~\cite{hobbes}. Invariant subgroups are encased in diamond shaped boxes; others are in circles.}
\label{fig:A4_graph}
\end{figure*}
Any of the vacuum alignments presented above automatically leads to a
non-trivial residual symmetry in the final Lagrangian. Actually, because all
elements of the triplet irrep of $\Delta(27)$ have eigenvalues $1,\, \omega$
or $\omega^2$, any pseudo-invariant vacuum alignment can be seen, with the
help of $h$, as an invariant one of another element of $\Delta(27)$. Notice
also that $h$ leaves pseudo-invariant any type of vacuum alignment. However,
it is not of real interest since $h \in \mathbb{Z}_3$, i.e. is the center of
$SU(3)$. As already pointed out, these transformations are included in the
$U(1)_Y$ part of full flavour group $G$. Therefore, we just need to look at
the $\Delta(27)/\mathbb{Z}_3$ group. In this case, some alignments in
Eq.~\eqref{nvevD27} will break the flavour group, but the full flavour group
will still be broken to a non-trivial subgroup.

The group-theoretical method presented above allows us to exclude several alignments without the need of dealing with the details of the scalar potential. However, this method tell us nothing about the global minima of the potential. Minimizing the general scalar potential for 3HDM is a very hard task, and until now no good method has been developed. When the potential exhibits some symmetry, the geometrical method~\cite{Degee:2012sk} can give us the answer (specially for large symmetry groups). This method has been used in the study of $\Delta(27)$ in 3HDM, where four classes of global minima are found~\cite{IgorCelso}, corresponding precisely to the eigenvectors given in Eq.~\eqref{nvevD27}.\footnote{We do not include minima which differ by an (irrelevant) overall phase. For example, one could have $(\omega, \omega^2, 1)$, but this minimum equals the phase $\omega$ multiplied by $(1, \omega, \omega^2)$, which is already taken into account.} Therefore, for the flavour group $\Delta(27)$ all the global minima of the potential are excluded by the theorem. Said otherwise, one cannot construct a viable model based on $\Delta(27)$ where $\Phi$ is in a faithful (triplet) irrep.

\subsection{The flavour group $A_4$ in 3HDM}
\label{sec:A4}

The flavour group $A_4$ can be viewed as the group defined by two generators
$s$ and $t$, with presentation
\begin{equation}
s^2=t^3=(st)^3=1.
\end{equation}
The group $A_4$ has one non-trivial invariant subgroup; $Z_2\times Z_2$, generated by $s$ and $tst^2$. It also has four subgroups $Z_3$, generated by $t$, $tst$, $st$ and $st^2$, and three subgroups $Z_2$ generated by $s$, $t^2st$ and $tst^2$. A summary of the subgroups and their generators is shown in Fig.~\ref{fig:A4_graph}, based on Ref.~\cite{hobbes}.

A possible matrix representation for the three-di\-men\-sion\-al irrep is
\begin{equation}
s=
\begin{pmatrix}
1&0&0\\
0&-1&0\\
0&0&-1
\end{pmatrix}\,,\quad
t=
\begin{pmatrix}
0&1&0\\
0&0&1\\
1&0&0
\end{pmatrix}\,.
\label{A4:s_and_t}
\end{equation}
We can use the group-theoretical method in order to find the set of
alignments forbidden by the theorem. In this case, the eigenvectors $(v_1,
v_2, v_3)$ associated to the generators of the subgroups are
\begin{eqnarray}
\begin{split}
t:&&\ (1,1,1),\ \
(1,\omega,\omega^2),\ \ (1,\omega^2,\omega);\\
tst:&&\ (-1,1,1),\ \ (-1,\omega,\omega^2),\ \ (-1,\omega^2,\omega);\\
st:&&\ (1,-1,1),\ \ (1,-\omega,\omega^2),\ \ (1,-\omega^2,\omega);\\
st^2:&&\ (1,1,-1),\ \ (1,\omega,-\omega^2),\ \ (1,\omega^2,-\omega);\\
s:&&\ (1,0,0),\ \ (0,v_2,v_3);\\
t^2st:&&\ (0,1,0),\ \ (v_1,0,v_3);\\
tst^2:&&\ (0,0,1),\ \ (v_1,v_2,0),
\end{split}
\label{nvevA4}
\end{eqnarray}
where $v_1, v_2$ and $v_3$ are complex parameters. Using the geometrical
method, the global minima of the $A_4$ symmetric scalar potential were found
to be~\cite{Degee:2012sk}
\begin{equation}
(1,1,1)\,,\, (1,0,0)\,,\,(\pm 1,e^{i \pi/3},e^{-i \pi/3})\,,
\,(1,e^{i\alpha},0)\,, \end{equation}
up to permutations and global phase transformations. All these global minima are included in the list of alignments given in Eq.~\eqref{nvevA4}. Indeed, the first two are obviously there; the third vev alignment is $(\pm 1,e^{i \pi/3},e^{-i \pi/3})$ $=-(\mp 1,\omega^2,\omega)$ and is in the list as well. Finally, the last minimum is a particular case when $|v_1|=|v_2|$. In summary, each $A_4$ global minimum leaves invariant a non-trivial subgroup of the full flavour symmetry $G=A_4\times U(1)_Y$ and is thus excluded by the no-go theorem. That is, one cannot construct a viable model based on $A_4$ where $\Phi$ is in a faithful (triplet) irrep~\cite{Felipe:2013ie}.

\begin{figure*}[t]
\centering
\includegraphics[width=11cm]{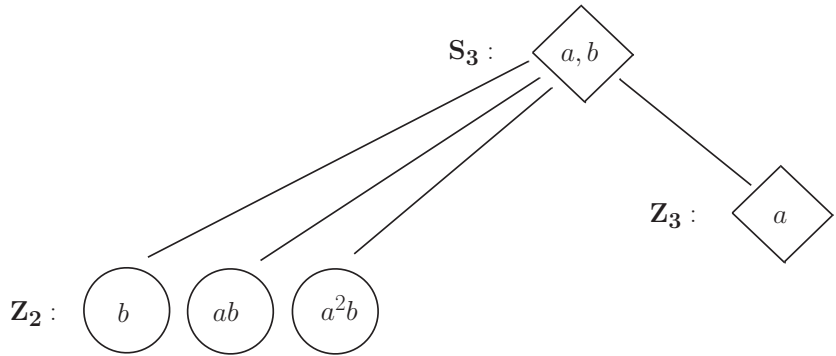}
\caption{Diagram of the subgroups of $S_3$ and their generators, based on Ref.~\cite{hobbes}. Invariant subgroups are encased in diamond shaped boxes; others are in circles.}
\label{fig:S3_graph}
\end{figure*}

It is interesting to consider an example similar to the one we included in point (ii) of Sect.~\ref{subsec:examples}, where the representations are $Q_L\sim (\boldsymbol{1},\boldsymbol{1}',\boldsymbol{1}'')$ and $\Phi\sim d_R\sim u_R\sim \boldsymbol{3}$. Now we take $\left<\Phi\right> \sim(1, e^{i \alpha}, 0)$. This leaves invariant a residual subgroup with generators $\mathcal{G}_L = \textrm{diag}(1,1,1)$ and $\mathcal{G}_{R}^{u} = \mathcal{G}_{R}^{d} = \mathcal{G} = \textrm{diag}(1,1,-1)$. The mass matrices for the up- and down-quarks may be written as
\begin{equation} \left( \begin{array}{ccc} a  &  a\, e^{i\alpha} & 0\\
 b & b\, \omega\, e^{i\alpha} & 0\\
c & c\, \omega^2\, e^{i\alpha} &   0 \end{array} \right). \end{equation}
This implies that, after up- and down-quark mass matrix diagonalisation, there remain three mixing angles and three non-degenerate quark masses. However, these matrices are not invertible, which signals the presence of massless quarks in both up and down sectors, in contradiction with the experiment. If we had looked exclusively at the left-handed quark sector, as in Ref.~\cite{Leurer:1992wg}, we would have accepted this model as a viable one. Although the symmetry has been completely broken in the left sector, the full Lagrangian has an unbroken symmetry in the right-handed quark sector. It is precisely here where the residual symmetry acts non-trivially and the no-go theorem applies.

\subsection{The flavour group $S_3$ in 3HDM}
\label{sec:S3}

Since the group $S_3$ is the smallest non-Abelian discrete symmetry, it has been widely used for flavour physics in the quark and lepton sectors (for recent works see e.g. Refs.~\cite{Ma:2013zca,Canales:2013cga}, and references therein). The flavour group $S_3$ can be viewed as the group defined by two generators, $a$ and $b$, with presentation
\begin{equation}
a^3=b^2=(ab)^2=1\,.
\end{equation}
It has the non-trivial invariant subgroup $Z_3$, which is generated by $a$, and three subgroups $Z_2$, generated by $b$, $ab$ and $a^2b$. We summarise in Fig.~\ref{fig:S3_graph} these subgroups and their generators~\cite{hobbes}.

The $S_3$ flavour group has no three-dimensional irrep.  Therefore, any 3HDM with a scalar potential invariant under this group, with a faithful representation, must be built from a two-dimensional irrep $\mathbf{2}$ and one of the one-dimensional irreps, $\mathbf{1}$ or $\mathbf{1}^\prime$. The representation of the generators for these irreps can be chosen as
\begin{align}
\begin{split}
\mathbf{2}:&\quad
a=\begin{pmatrix}
\omega&0\\
0&\omega^2
\end{pmatrix}\,,\,
b=\begin{pmatrix}
0&1\\
1&0
\end{pmatrix},\\
\mathbf{1}:&\quad a=b=1,\\
\mathbf{1}^\prime:&\quad a=-b=1.
\end{split}
\end{align}
When building an $S_3$ flavour model we have to decide whether our reducible
three-dimensional representation is $\mathbf{2}+\mathbf{1}$ or
$\mathbf{2}+\mathbf{1}^\prime$. We may represent them in the compact form
\begin{equation}
\mathbf{3}_{\pm}:\quad
a=
\begin{pmatrix}
\omega&0&0\\
0&\omega^2&0\\
0&0&1
\end{pmatrix}\,,\quad
b_{\pm}=
\begin{pmatrix}
0&1&0\\
1&0&0\\
0&0&\pm 1
\end{pmatrix}\,,
\label{S3:a_and_b}
\end{equation}
with the plus and minus signs assigned to $\mathbf{2}+\mathbf{1}$ and $\mathbf{2}+\mathbf{1}^\prime$, respectively.  Following the group-theoretical method, we can find the set of vev alignments forbidden by the theorem. In this case, the eigenvectors associated to the reducible generators of the subgroups will be denoted by $((v_1,v_2),u)$, in order to distinguish the doublet from the singlet irrep. The corresponding eigenvectors for the $\mathbf{\mathbf{3}}_{\pm}$ are
\begin{eqnarray}
\begin{split}
a:&&\ ((1,0),0),\ ((0,1),0),\ ((0,0),1) ;\\
b_+:&&\ ((v,v),u),\ ((-1,1),0) ;\\
b_-:&&\ ((1,1),0),\ ((-v,v),u) ;\\
ab_+:&&\ ((v\,\omega,v),u),\ ((-\omega,1),0) ;\\
ab_-:&&\ ((\omega,1),0),\ ((-v\,\omega,v),u) ;\\
a^2b_+:&&\ ((v\,\omega^2,v),u),\ ((1,-1),0) ;\\
a^2b_-:&&\ ((\omega^2,1),0),\ ((-v\,\omega^2,v),u) .
\end{split}
\label{nvevS3}
\end{eqnarray}
Thus, if the $S_3$ singlet $\mathbf{1}$ acquires a vev, we can then conclude
from Eqs.~\eqref{nvevS3} that we are not allowed to have the vev alignments
$(1,1)\,,\,(\omega,1)\,,\,(\omega^2,1)$ for the two Higgs doublets in the
$S_3$ doublet irrep. Similarly, if the $S_3$ singlet $\mathbf{1}^\prime$
acquires a vev, the alignments $(-1,1)$, $(-\omega,1)$, $(-\omega^2,1)$ are
excluded. Of course, it remains to be seen whether all global minima of the
$S_3$-invariant Higgs scalar potential are contained or not in the set of
forbidden vev alignments.

\section{Conclusions}
\label{sec:conclusions}

In this work we have studied the connection between the breaking of flavour
symmetries by the Higgs vevs and the existence of residual symmetries in the
quark mass matrices. We have performed two tasks. First, in order to avoid
nonphysical quark masses and mixing, we have developed a simple but powerful
no-go theorem, highlighting the importance of including the transformation
properties of the right-handed fields. Second, we have shown that, in many
instances, exploring the eigenspaces of all the subgroups of the original
flavour group is sufficient to study all the relevant vevs. The vevs of the
scalars coupled to the quarks have to break the full flavour symmetry (or
break it into a subgroup acting as baryon number) in order to avoid
nonphysical quark masses and mixing.

In this context, the notion of full flavour group turns out to be important. While, in general, one refers to the flavour group as the group added to the SM acting globally on the flavour space, the SM gauge group already contains the global hypercharge transformation that should also be taken into account. The inclusion of this additional flavour transformation builds what we call the full flavour group. Then the problem of finding the vevs left invariant by the full flavour group and, therefore, excluded by the theorem, turns into the problem of finding all pseudo-invariant vevs of the flavour group, discussed here for the first time.

As we have shown through some examples, it is possible to find a set of excluded vev alignments (minima or not of the scalar potential) just by determining the eigenvectors of the Higgs representation for each flavour group element. If the global minima are known, as it is the case of 3HDM with $A_4$ or $\Delta(27)$ symmetric potentials, one then needs only to check whether these minima are contained in the set of excluded alignments. For 3HDM with $A_4$ (or $S_4$) and $\Delta(27)$ symmetric potentials, this is indeed verified and thus these models are excluded. In this case, a phenomenologically viable description of the quark sector requires that (1) the symmetry is explicitly broken by new interaction terms, or (2) higher-order interaction terms in the Higgs potential are present, which could then lead to the complete breaking of the $A_4$ or $\Delta(27)$ group upon minimisation of the potentials, or (3) additional non-invariant scalar multiplets are added to the theory. In more realistic but non-minimal models, one or more of these options are necessarily implemented. The same type of analysis can be carried out for other groups and, in particular, for smaller groups which have a more complex scalar potential.

While we only treated quarks in our discussion, the extension of the theorem to the whole fermion sector, including leptons, is straightforward. If neutrinos are Dirac-type particles the analogy is direct. There is the experimental possibility of a massless neutrino, however, as shown in our analysis, all cases with CP violation always imply a massless fermion in each sector. The case where neutrinos are Majorana-type particles is more interesting since there is a larger number of possible implementations. Lepton number is no longer conserved and thus the theorem will be slightly modified.

\noindent\emph{Note added:} While our paper was undergoing the review process, Ref.~\cite{Fonseca:2014koa} appeared, in which a classification of lepton mixing matrices is performed based on the assumption that the residual symmetries in the charged-lepton and neutrino mass matrices originate from a finite flavour symmetry group.

\begin{acknowledgements}
J.P.S. is grateful to John Jones for clarifications about the notation in Ref.~\cite{hobbes}. We are grateful to Yossi Nir for enlightening discussions regarding Ref.~\cite{Leurer:1992wg}. The work of R.G.F. and J.P.S. was partially supported by FCT - \textit{Funda\c{c}\~{a}o para a Ci\^{e}ncia e a Tecnologia}, under the projects PEst-OE/FIS/UI0777/2013 and CERN/FP/123580/2011, and by the EU RTN Marie Curie Project PITN-GA-2009-237920. The work of C.C.N. was partially supported by Brazilian CNPq and Fapesp. The work of I.P.I. is supported by the RF President grant for scientific schools NSc-3802.2012.2, and the Program of Department of Physics SC RAS and SB RAS ``Studies of Higgs boson and exotic particles at LHC". The work of H.S. is funded by the European FEDER, Spanish MINECO, under the grant FPA2011-23596, and the Portuguese FCT project PTDC/FIS-NUC/0548/2012.
\end{acknowledgements}

\appendix

\section{Linear independence of Yukawa matrices}
\label{app:proof}

The proposition presented in Sect.~\ref{subsec:group} relies on the statement that if a set of Higgs doublets $\phi_k$ realises an irreducible representation of the flavour symmetry group $G$, then either all of them decouple from all up-quarks or all down-quarks, or none of them can decouple from them. This statement is stronger than just saying that none of the matrices $\Gamma^k$ or $\Delta^k$ can be zero in a given Higgs basis. It says that this cannot happen in any Higgs basis. The basis-invariant formulation of this requirement is that $\Gamma^k$ or $\Delta^k$ are either all zeros or are linearly independent.

In this Appendix, we provide an accurate proof of this statement. We start
with the following proposition:
\newline\newline
{\bf Proposition} \emph{Let $\{\Gamma^k\}$, $k = 1,\dots,N$, be a
    non-zero element of a complex vector space $\bigotimes_N V_\Gamma$;
    $V_\Gamma$ is the space of each $\Gamma^k$. Similarly, let
    $\{\phi_k\}$, with $k = 1,\dots,N$, be elements of a complex vector
    space $\bigotimes_N V_\phi$, where $V_\phi$ is the space of each
    $\phi_k$. Suppose that a symmetry group $G$ acts in the space of
    $\{\phi_k\}$, and $\{\phi_k\}$ realise an irrep of $G$. Finally,
    suppose that there exists an expression ${\cal L}$, proportional to
    $\Gamma^k \phi_k$ (summation over repeated indices assumed), which is
    invariant under $G$. Then, $\Gamma^k$ are linearly independent.}
\newline

Before proceeding to the proof, a few remarks are in order. First, to make the notation a bit more familiar, we notice that in our case $V_\Gamma = \mathbb{C}^9$ of complex-valued $3\times 3$ matrices $\Gamma^k$, and $V_\phi = \mathbb{C}^2$, the space of Higgs doublets. However, the proposition itself is not specific to this particular space. Also, we note that the linear independence among $\Gamma^k$ is a basis-invariant way of saying that no individual $\Gamma^k$ will ever become zero after any basis change in $\phi_k$. Finally, $\{\Gamma^k\}$ being a non-zero element of $\bigotimes_N V_\Gamma$ means that not all individual $\Gamma^k=0$.
\newline\newline
\noindent\emph{Proof} We shall prove the statement by contradiction: assuming that $\Gamma^k$ are linearly dependent, we will show that none $\{\phi_k\}$ can be an irrep.

Suppose that $\Gamma^k$ are linearly dependent, namely, that there exist
complex coefficients $c_k$, not all of them being zeros, such that $c_1
\Gamma^1 + \cdots + c_N  \Gamma^N = 0$. Denote the number of linearly
independent $\Gamma^k$ by $p$, with $0 < p < N$. Then, it is possible to
perform a correlated basis transformation in the spaces of $\Gamma$'s and
$\phi$'s leaving invariant $\Gamma^k \phi_k$, which sets $N-p$ matrices
$\Gamma$'s to zero:
\begin{eqnarray}
&&\mbox{for}\ k = 1,\dots,p:\, \Gamma^k \not = 0,
\mbox{$\Gamma^k$ are linearly independent,} \nonumber\\
&&\mbox{for}\ k = p+1,\dots,N:\, \Gamma^k = 0\,. \label{split}
\end{eqnarray}
In this basis, we call the corresponding Higgs doublets $\phi_k$ {\em active} for $k = 1,\dots,p$ and {\em passive} for $k = p+1,\dots,N$. The expression ${\cal L}$ is thus written only in terms of active fields. Let us now apply a transformation $g \in G$, which leads to
\begin{equation}
\Gamma^k \phi_k \mapsto \Gamma^k \phi^g_k = \Gamma^k ({\cal G}_g)_{kl}
\phi_l = \Gamma^l ({\cal G}_g)_{lk} \phi_k \,.
\label{Gammak:eq1}
\end{equation}
The last transformation here is just a relabeling of the indices. Note that we do not require that $\Gamma^k \phi_k$ itself is invariant under $g$, because ${\cal L}$ can contain other overall factors which are also transformed under $g$ (in our particular case, these are left- and right-handed quark fields). These additional factors lead to the appearance of extra transformations (such as ${\cal G}_L$ and ${\cal G}_R^d$), which are unitary transformations of the space $V_\Gamma$ and transform each individual $\Gamma^k$, but do not mix different $\Gamma$'s. In particular, they cannot make a zero $\Gamma^k$ non-zero, and vice versa, nor can they convert a linearly independent set of $\Gamma$'s into a linearly dependent one.

By construction, the superscript $l$ in the last expression of Eq.~(\ref{Gammak:eq1}) refers to active fields. Since $\Gamma^l ({\cal G}_g)_{lk}$ is a linear combination of linearly independent $\Gamma$'s, if $k$ also corresponds to an active field, then this linear combination can stay non-zero, and $({\cal G}_g)_{lk}$ can be anything. If $k$ corresponds to a passive field, then this linear combination is zero, which can happen only when $({\cal G}_g)_{lk} = 0$ for all active $l$'s. Since ${\cal G}_g$ is unitary, it also follows that $({\cal G}_g)_{lk} = 0$ for all passive $l$ and active $k$. In other words, the group transformation ${\cal G}_g$ does not mix active and passive fields, and hence it has a block-diagonal form in the space of $\phi$'s. Repeating this analysis for all $g \in G$, with the same conclusion, we find that active and passive fields represent two invariant subspaces of the space $\{\phi_k\}$. Thus, the representation is reducible. This completes the proof.

The above proposition can be immediately applied to the one given in
Sect.~\ref{subsec:group}. Suppose that we pick up Higgs doublets which
realise an irrep of the flavour group $G$; this can be either the entire set
of doublets of the model or, in the case when the Higgs doublets are in a
reducible representation, this can be an irreducible subspace of the Higgs
fields. In any of these cases, we apply the above proposition, assuming that
$N$ stands for the dimension of the chosen irreducible representation. With
this minor modification, we obtain exactly the proposition from the main
text.

\section{Block-diagonal CKM matrix}
\label{block.diag}

We prove here the assertion made after Eq.~\eqref{symHuHd}, that if the
generator $\mathcal{G}_{Lg_1}$ given in Eq.~\eqref{symHuHd} is not
proportional to the identity matrix, then the CKM matrix is block-diagonal.

The matrix $\mathcal{G}_{Lg_1}$, being a $3\times 3$ unitary matrix, is diagonalizable and we can find three orthonormal eigenvectors. If $\mathcal{G}_{Lg_1}$ is not proportional to the identity, then there is at least one non-degenerate eigenvalue $e^{i\theta}$. Take its associated eigenvector $\mathbf{w}$, for which $\mathcal{G}_{Lg_1}\mathbf{w}=e^{i\theta}\mathbf{w}$. It follows from Eq.~\eqref{symHuHd} that
\begin{equation}
\mathcal{G}_{Lg_1}(H_d\,\mathbf{w})=H_d\,\mathcal{G}_{Lg_1}\mathbf{w}
=e^{i\theta}(H_d\,\mathbf{w})\,,
\end{equation}
i.e., $(H_d\,\mathbf{w})\neq\boldsymbol{0}$ is also an eigenvector of $\mathcal{G}_{Lg_1}$ with eigenvalue $e^{i\theta}$. Since this eigenvalue is non-degenerate, $(H_d\,\mathbf{w})$ should be proportional to $\mathbf{w}$: $H_d\,\mathbf{w}=m^2_{d_i}\mathbf{w}$, where obviously $m_{d_i}$ is one of the down-quark masses. Hence, $\mathbf{w}$ fills one of the columns of the matrix $V_d$ which diagonalises $H_d$. Similar arguments make the same vector $\mathbf{w}$ to be in one of the columns of the matrix $V_u$, which diagonalises $H_u$. The CKM matrix $V=V_u^\dag V_d$ then contains one row and one column consisting of zeros, except for a unit entry where they cross. Finally, the vector $\mathbf{w}$ will remain as a common eigenvector of $H_d$ and $H_u$ even if $H_d\mathbf{w}=\boldsymbol{0}$ or $H_u\mathbf{w}=\boldsymbol{0}$ (some massless quark). This proves the assertion.

\section{Yukawa structure}
\label{ap:yukawa}

We show here that some combination of representations cannot be assigned to the fields $\bar{Q}_L,\Phi,d_R$ and $u_R$ in order to construct group invariants in the Yukawa interactions of Eq.~\eqref{LY_before}. For instance, it is not possible that only one of $\bar{Q}_L,\Phi,d_R$ or $u_R$ is assigned to a faithful irrep or contains a faithful irrep.

We begin by reviewing the following result: given two vectors (fields) $x$ and $y$ that transform under irreducible representations $\mu$ and $\nu$ of $G$, the unique invariant present in $x \times y$ is $x \cdot y=x_i\,y_i$, only possible when $\mu=\nu^*$. The proof relies on the Schur lemmas. The invariant in $x\times y$ can be written as $I=x_i\,y_j\,C_{ij}$ where $C_{ij}$ are the Clebsch-Gordan coefficients that connect $x\times y$ to the trivial representation. Invariance of $I$ requires
\begin{equation}
(D^{(\mu)}(g))^TC D^{(\nu)}(g)=C,\quad \mbox{for all $g$ in $G$.}
\end{equation}
We can rewrite it as $C D^{(\nu)}(g)=(D^{(\mu)}(g))^*C$. One of the Schur
lemmas requires that $C\neq \boldsymbol{0}$ only if $\mu\simeq\nu^*$.
However, if $\mu=\nu^*$, another Schur lemma ensures that $C=\mathbbm{1}$.
Now, this result also applies if $x$ and $y$ are reducible and composed of
more than one irreducible piece: complex conjugate irreps need to be
contracted to build invariants.

Let us now analyse $\bar{Q}_L\times\Phi\times d_R$. Suppose that $Q_L$ and
$\Phi$ are assigned to unfaithful representations and $d_R$ to a faithful
irrep. Then the representation of $\bar{Q}_L\times\Phi$ is also unfaithful
and does not contain any faithful representation that can be contracted to
$d_R$.

\end{document}